\documentstyle[preprint,aps]{revtex}
\begin{document}

\draft

\preprint{$
\begin{array}{l}
\mbox{UCD--96--07}\\[-3mm]
\mbox{AMES-HET-96-01}\\[-3mm]
\mbox{March 1996}\\
\end{array}
$}
\title{Top-Quark Decay Via the Anomalous Coupling $\bar t c \gamma$ 
\\ at Hadron Colliders}

\author{T. Han$^a$, K. Whisnant$^b$, B.-L. Young$^{b,c}$ and X. Zhang$^b$ }

\address{{ $^a$Department of Physics, University of California, Davis,
 CA 95616, USA}\\
{$^b$Department of Physics and Astronomy,
 Iowa State University, Ames, IA 50011, USA}\\
{$^c$Institute of Physics, Academia Sinica, Taipei, Taiwan}\\
}
\maketitle

\begin{abstract}

We determine the constraints on the anomalous top-quark coupling
associated with the flavor-changing neutral current vertex 
$\bar t c \gamma $ from the limits on the $b$-quark rare decay $b\rightarrow
s \gamma$ and non-standard top-quark decays. Based on these constraints, we 
discuss the experimental observability of the rare decay mode $t \rightarrow 
c \gamma $, both at the Fermilab Tevatron with a 
luminosity-upgrade and at the LHC.

\end{abstract}
              
%\pacs{??}

\narrowtext

\section{Introduction}

The top quark has been experimentally observed at the Fermilab
Tevatron by the CDF and D0 collaborations \cite{CDFD0} with a measured
mass around 175 GeV. Because it has a mass of the order of the Fermi
scale, the top quark couples quite strongly to the electroweak
symmetry-breaking sector.
%and to the longitudinal components of the gauge bosons.
In the Standard Model (SM), the electroweak symmetry-breaking sector 
consists of a complex
fundamental Higgs scalar, but ``triviality'' \cite{tri} and
``naturalness'' \cite{natural} of the scalar sector suggest that the
Higgs sector may not be so simple. It is therefore plausible to assume
that the mass generation mechanism is more complicated and the Higgs
sector of the SM is just
an effective theory, and that new physics phenomena may be
manifested through effective interactions of the top quark \cite{PZ}.

If anomalous top-quark couplings beyond the SM exist, they will affect
top-quark production and decay processes at hadron and $e^+e^-$
colliders \cite{cmy,akr}.  Furthermore, such couplings would certainly
affect certain low-energy quantities which are measured with high
precision. One such possibility is the partial width ratio
$R_b = \Gamma(Z \rightarrow b \bar b) / \Gamma(Z\rightarrow {\rm Hadrons} )$
measured at LEP-I, which is about 3$\sigma$ higher than the Standard Model
expectation \cite{LEP} and could be an indication of new physics
associated with the heavy top quark.

In Ref.~\cite{HPZ}, the experimental constraints on an anomalous top-quark 
coupling $Z \bar t c$ and the experimental observability of the induced rare 
decay mode $t \rightarrow Z c$, at the Fermilab Tevatron and the
LHC at CERN, have been investigated in detail. In this paper, we examine
the possible anomalous top-quark coupling $\bar t c \gamma$ and its
implications in both low and high energy processes. In our analysis
we will also allow for the possibility of a $\bar t c g$ coupling
because it has a similar form to the $\bar t c \gamma$ coupling. 
Since the Standard Model prediction of $\Gamma(t\rightarrow c\gamma)$
is unobservably small \cite{tcgammaSM}, any experimental evidence for
the $\bar t c \gamma$ coupling will be an unequivocal indication of new
physics beyond the Standard Model.

This paper is organized as follows. In Section II, we examine the
constraints on the anomalous $\bar t c \gamma (g)$ couplings from
$b \rightarrow s \gamma$ and from the limit on non-standard top decays
at the Tevatron. In Section III, we study the possibility of detecting
the $\bar t c \gamma$ coupling at the Tevatron and the LHC, paying
particular attention to the kinematical characteristics to extract the
signal from the possible background. The cuts are adjusted to the
different physics requirements at the two accelerators. Finally,
in Section IV, we summarize our results.

%\section{ constraints on the anomalous  $ {\bar t}  c \gamma (g)$
%couplings from $b \rightarrow s \gamma$   }

\section{ constraints on the top-quark anomalous couplings }

Following Ref.~\cite{PZ}, we introduce an effective Lagrangian involving the 
anomalous top-quark couplings,
\begin{equation}
{\cal L}^{eff} = {\cal L}^{SM} + \Delta {\cal L}^{eff}
\,
\end{equation}
where ${\cal L}^{SM}$ is the Standard Model Lagrangian and $\Delta {\cal
L}^{eff}$ includes the anomalous top-quark couplings. For the purpose of this
paper, we consider only the lowest dimension, $CP$-conserving operators which
give rise to anomalous $ \bar t c \gamma (g)$ vertices. Then
\begin{eqnarray}
\Delta {\cal L}^{eff} =    \frac{1 }{ \Lambda } \,
[ \kappa_{\gamma} e   \bar t \sigma_{\mu\nu} c
F^{\mu\nu} + \kappa_g g_s \bar t
\sigma_{\mu\nu} \frac{\lambda^i}{2} c G^{i \mu\nu}]
+ {\it h.c.} \ ,
\label{Leff}
\end{eqnarray}   
where $F^{\mu\nu}$ ($G^{i \mu\nu}$) is the $U_{em}(1)$ $( SU_c(3))$ field 
strength tensor; $e$ ($g_s$) is the corresponding coupling constant;
%of the $U_{em}(1)$ ($SU_c(3)$); 
and $\Lambda$ is the cutoff of the effective theory, which
we take to be the order of 1 TeV. The parameters $\kappa_\gamma$ and
$\kappa_g$ can be interpreted as the strengths of the anomalous
interactions, or, alternatively, $\Lambda/\kappa$ is the approximate
scale at which new physics in the top-quark sector occurs.

%In this latter
%interpretation the new physics scale for the anomalous 
%chromo- and electromagnetic couplings are different,
%characterized by $\kappa_g$ and $\kappa_\gamma$.

The measurement of the inclusive branching ratio for the process $b
\rightarrow s \gamma$ \cite{cleo} puts severe constraints on various
extensions of the standard model \cite{hewett}. Here we study its
constraints on anomalous top-quark couplings $\bar t c \gamma$ and
$\bar t c g$. Anomalous couplings of various types associated with
the top quark and their
constraints obtained from low-energy processes have been considered
in the literature \cite{HRFY}.

%Our
%approach is somewhat parallel to that of Hewett and Rizzo \cite{HR} in a
%discussion of anomalous flavor conserving top-quark couplings, and that
%of Fujikawa and Yamada\cite{FY} in a discussion of an anomalous $W\bar t
%b$ coupling constrained by $b \rightarrow s \gamma$.

The effective Hamiltonian for $b \rightarrow s \gamma$ is given 
by \cite{GSW,CFM,Buras},
 \begin{equation}
H_{eff} = - \frac{4}{\sqrt 2} G_F V^*_{ts}V_{tb} \sum_i C_i(\mu)
                 O_i(\mu)   ,
\end{equation}
where $i$ runs from 1 to 8, and $\mu$ denotes the energy scale at which
$H_{eff}$ is applied.  The operators set $O_i( \mu )$ consist of six 
four-quark operators $O_{1-6}$, the electromagnetic dipole operator $O_7$,
and the chromo-magnetic dipole operator $O_8$. At the weak scale, the only 
contributing operator is $O_7$. However, when evolving down to the low energy 
scale $\mu \sim m_b$, $O_7$ will mix with operator $O_8$ and others. Here, 
$V_{ij}$ are the elements of the Cabbibo-Kobayashi-Maskawa mixing matrix.

The partial decay width for $b \rightarrow s \gamma$ is, neglecting the 
strange-quark mass, 
\begin{equation}
\Gamma( b \rightarrow s \gamma )= \frac{ \alpha G_F^2 m_b^5}{128 \pi^4}
{|V_{ts}^*V_{tb}C_7(m_b)|}^2 .
\end{equation}
Normalized by the inclusive semileptonic branching ratio, the $b \rightarrow s
\gamma$ branching ratio can be written as
\begin{equation}
\frac{ BR( b \rightarrow s \gamma )}{BR(b \rightarrow c e \nu )}=
\frac{ {|V_{ts}^*V_{tb} |}^2 }{ {| V_{cb} |}^2 } \frac{6 \alpha}{\pi
g(z)} {| C_7(m_b) |}^2 ,
\end{equation}
where $g(z) = 1-8 z^2 + 8z^6 -z^8 -24z^4 \ln z$, with
$z= m_c/ m_b$. In the standard model, the complete leading logarithmic
approximation gives \cite{Buras}
\begin{equation}
C_7(m_b) = \eta^{16/23} C_7(M_W) + \frac{8}{3}( \eta^{14/23} - \eta^{16/23} )
      C_8(M_W) + C_2(M_W)\sum_{i=1}^8 h_i \eta^{p^i}   ,
\end{equation}
with $\eta= \alpha_s(M_W) / \alpha_{s}(m_b)$ . The coefficients
$C_2(M_W)$, $h_i$ and $p^i$ can be found in~\cite{Buras}.

The anomalous top-quark couplings $\bar t c \gamma$ and $\bar t c g$
modify the coefficients of operators $O_7$ and $O_8$. The coefficients
of these operators at the electroweak scale can be written as
\begin{equation}
C_7(M_W) = C_7^{SM}(M_W) -\frac{1}{2} {( \frac{V_{cs}}{V_{ts}} )}^*
\frac{m_t}{\Lambda} \ln( \frac{\Lambda^2}{m_t^2} ) \kappa_\gamma ,
\end{equation}

\begin{equation}
C_8(M_W) = C_8^{SM}(M_W) - \frac{1}{2} {( \frac{ V_{cs}}{V_{ts}} )}^*
\frac{m_t}{\Lambda} \ln( \frac{\Lambda^2}{m_t^2} ) \kappa_g,
\end{equation}
where $C_7^{SM}(M_W)$ and
$C_8^{SM}(M_W)$ are the standard model contributions.

Using $BR( b \rightarrow c e \nu ) = 0.108$ and
the recent CLEO measurement \cite{cleo}
\begin{equation}
1 \times 10^{-4} < BR( b \rightarrow s \gamma ) < 4.2 \times 10^{-4},
\label{EQ:BSGLIM}
\end{equation}
we plot the allowed region for $\kappa_\gamma$ and
$\kappa_g$ in Fig.~1, for $m_b=5$~GeV, $m_t = 175$ GeV,
$\alpha_s(M_Z) =0.12$ and 
$|V_{ts}^*V_{tb} / V_{cb} |^2=0.95$ \cite{Buras}.  
Throughout this paper, we take $\Lambda=1$ TeV.
In the figure there are two allowed bands:
the right band corresponds to a small
anomalous correction to the SM $C_7(m_b)$ coefficient which obeys the
limit in Eq.~(\ref{EQ:BSGLIM}), while the left band corresponds to a
large correction which is approximately twice the magnitude of the SM
contribution but of the opposite sign, so that the magnitude of
$C_7(m_b)$ is still consistent with Eq.~(\ref{EQ:BSGLIM}).

A further restriction on $\kappa_\gamma$ and $\kappa_g$ can be obtained from 
the fact that $t\rightarrow cg$ and $t\rightarrow c\gamma$ decays are not the 
most significant top decay modes as shown by the recent CDF data \cite{cdftbw}.
A straightforward calculation of partial widths yields
\begin{equation}
{\Gamma(t\rightarrow c\gamma)\over \Gamma(t\rightarrow bW)}
= {16\sqrt2\pi\alpha\over
G_F\left(1-{M_W^2\over m_t^2}\right)^2\left(1+{2M_W^2\over m_t^2}\right)}
\left({\kappa_\gamma\over\Lambda}\right)^2,
\end{equation}
where the masses of the $c$ and $b$ quarks have been ignored. For the
case of $\Gamma(t\rightarrow cg)$, $\alpha$ is replaced by $4\alpha_s/3$
in the above equation.

The CDF data on the branching ratio of top decaying to $b$ \cite{cdftbw},
\begin{equation}
BR(t\rightarrow bW)=0.87^{\pm0.13\pm0.13}_{\pm0.30\pm0.11}
\end{equation}
places the limit
\begin{equation}
BR(t\rightarrow cg) < 0.45
\end{equation}
at the one standard deviation level, where the insignificant contribution of
$t \rightarrow c\gamma$ has been ignored. This constraint gives the limit 
\begin{equation}
|\kappa_g|<0.9,
\label{EQ:KGLIM}
\end{equation}
and is shown in Fig.~1 as the dashed lines.

One can see that $BR( b \rightarrow s \gamma )$ combined with
$BR(t\rightarrow bW)$ has constrained $\kappa_\gamma$ very strongly.
The branching ratio for $t \rightarrow c \gamma$ is shown in Fig.~2(a)
versus $\kappa_\gamma$ for the maximum and minimum allowed magnitudes of
$\kappa_g$. Specifically, we find that
\begin{eqnarray}
{\rm for} \; \kappa_g=0: \; 
|\kappa_\gamma|&<&0.16 \quad {\rm and} \quad
BR(t \rightarrow c \gamma) < 1.3\times10^{-3} \\
{\rm for} \; |\kappa_g|=0.9: \; 
|\kappa_\gamma|&<&0.28 \quad {\rm and} \quad
BR(t \rightarrow c \gamma) < 2.2\times10^{-3},
\label{EQ:KAPPA}
\end{eqnarray}
when accounting
for the contribution of the $t\rightarrow cg$ decay to the total width.
A similar constraint for the case with $\kappa_g=0$ has been obtained
in Ref.~\cite{mich}.
The branching ratio for $t \rightarrow c g$ is shown
in Fig.~2(b) versus $\kappa_g$.
%the ratio $\Gamma(t \rightarrow c\gamma)/\Gamma(t \rightarrow bW)$
%has a maximum value of $4.0\times10^{-3}$, which corresponds to

%\section{Top-quark decay to $c \gamma (g)$ at hadron colliders}
\section{Top-quark decay via anomalous couplings at hadron colliders}

At the Fermilab Tevatron, the cross section for $t{\bar t}$ production
is about 5~pb at $\sqrt{s}=2$~TeV. A detailed study of top-quark decays
is possible with the expected 1~fb$^{-1}$/yr integrated luminosity of
the Main Injector, or the 10~fb$^{-1}$/yr anticipated at the Tevatron
Upgrade \cite{tevlum}. Top-quark production is much greater at the LHC,
with a cross section being of order 1000 pb, due to the
larger center of mass energy (14~TeV) and higher integrated luminosity
(100~fb$^{-1}$/yr). Therefore it is feasible to search for evidence
of anomalous couplings such as ${\bar t}c\gamma$ at hadron colliders.

The decay $t\rightarrow c\gamma$ has already
been discussed in the previous section. To obtain the signal event rate,
we calculate the top-quark production via
$q{\bar q},gg\rightarrow t{\bar t}$ with the lowest order matrix
elements, but normalize the total cross sections to values which
include order $\alpha_s^3$ corrections \cite{hocs}. We thus find 
that the constant $K$ factors are
$$ K=1.4 \quad {\rm for \ \ Tevatron}; \quad \quad
   K=2.0 \quad {\rm for \ \ LHC}.
$$
For the parton
distributions we use the recent parameterization MRS Set-A \cite{mrsa}.
Due to the enormous QCD backgrounds at hadron colliders, it is very
difficult, if not impossible, to search for the signal if both the top
and antitop decay purely hadronically. We therefore look for events with
one top-quark decay leptonically
$t \rightarrow W^+b \rightarrow \ell^+\nu b$
(or $\bar t \rightarrow W^-\bar b \rightarrow \ell^-\bar \nu \bar b$),
where $\ell=e,\mu$.
%the type $t{\bar t} \rightarrow Wbc\gamma \rightarrow b\ell^\pm\nu c\gamma$.
The cross section for such events is calculated using the exact matrix
elements for an on-shell $W$. We have ignored the spin correlations for
the decaying top quarks since the top-quark production mechanisms we
consider give insignificant top-quark polarization \cite{nopol}.

%\subsection{$t \bar t\to Wb c \gamma$ at Hadron Colliders}

Figure~3 shows the calculated total cross section plotted versus 
$\kappa_\gamma$, for two values of $\kappa_g$, for the process
$t{\bar t} \rightarrow bWc\gamma \rightarrow jj\ell^\pm\nu \gamma$,
where $\ell=e,\mu$ and $j$ represents a jet from either a $b$ or $c$
quark. The curves stop at the maximum allowed value of $\kappa_\gamma$
in each case. Figure~3(a) is for the Fermilab Tevatron
($p\bar p$ collisions with $\sqrt s = 2$ TeV), and Fig.~3(b) for the 
LHC ($p p$ collisions with $\sqrt s = 14$ TeV).
We see that for the allowed values of $\kappa_\gamma$, the
signal cross section at the Tevatron could be as large as 4~fb, which is
rather small but still might be observable at the Tevatron Upgrade.
At the LHC, the maximal signal cross section is nearly 600~fb, which
means that there could be a large number of events even for non-maximal
couplings. We should remark that the top-quark production cross section
is somewhat sensitive to the top-quark mass. For $m_t=200$~GeV, for
example, the cross section is reduced by slightly
more or less than 50\% at the Tevatron and the LHC, respectively.

The only irreducible backgrounds to the signal
are electroweak (EW) $W^\pm \gamma$ production plus two QCD jets
\begin{equation}
p{\bar p}\rightarrow W^\pm \gamma jj \rightarrow \ell^\pm \nu \gamma jj. 
\label{EQ:EW}
\end{equation}
There are also possible ``fake'' events 
\begin{equation}
p{\bar p}\rightarrow W^\pm jjj\rightarrow \ell^\pm \nu \gamma jj 
\label{EQ:QCD}
\end{equation}
where one of the QCD jets fakes a photon. These backgrounds
have substantial production rates. The signal is however distinctive: 
besides the  isolated charged lepton and large missing transverse 
energy ($E_T^{miss}$) from $W^\pm$ decay, there is
a highly energetic photon, and two hard ($b,c$-quark) jets with which 
$c\gamma$ and $bW$ both reconstruct a top (or $\bar t$). Without
requiring $b$-tagging, our signal selection procedure is as follows:
we first examine the two values of the invariant mass $M(\gamma j_1)$
and $M(\gamma j_2)$ and identify the jet which gives an $M(\gamma j)$
value closer to $m_t$ as the $c$-quark jet. Naturally, the other jet
will be identified as $b$-quark jet. Due to the missing neutrino from 
the $W$ decay, the $W$ momentum can't be reconstructed unambiguously
due to the lack of knowledge of the parton c.m. frame.
This is known as the two-fold ambiguity in constructing the neutrino 
momentum along the beam direction \cite{pnu}.
Taking the $W$ mass as an input and for massless
leptons,  using the measured charged lepton momentum ($p^\ell$)
and the {\it transverse} momentum (with respect to the beam direction)
of the neutrino ($p_T^\nu$), the two solutions for the
{\it longitudinal} momentum of the neutrino can be expressed by
\begin{eqnarray}
\noalign{\vskip 5pt}
p_L^{\nu} = {1\over 2\, (p_T^{\ell})^2 } \Biggl\{ p_L^{\ell} 
\Bigl(M_W^2 + 2\, \hbox{\bf p}_T^{\ell} \cdot \hbox{\bf p}_T^{\nu} \Bigr)  
\pm \, p^\ell \, \biggl[ \Bigl(M_W^2 + 2\, \hbox{\bf p}_T^{\ell} \cdot 
\hbox{\bf p}_T^{\nu} \Bigr)^2 
- 4\, (p_T^\ell)^2\, (p_T^\nu)^2 \biggr]^{1/2} \Biggr\} \>.
\label{plnu}
\end{eqnarray}
There are therefore two solutions for $p_W^{}$, correspondingly. 
We again take the one that gives an $M(bW)$ value closer to $m_t$.
With this procedure, both top-quark momenta are experimentally
identifiable.

To make the calculation more realistic, we 
simulate the detector effects by assuming a Gaussian energy smearing
for the electromagnetic and hadronic calorimetry as follows:
\begin{eqnarray}
\Delta E/E &=& 30\%/\sqrt E \oplus 1\%, \quad  
{\rm for \ \ lepton \ and \ photon} \nonumber\\
&=&  80\%/\sqrt E \oplus 5\%, \quad  {\rm for \ \  jets,} 
\label{smear}
\end{eqnarray}
where the $\oplus$ indicates that the $E$-dependent and $E$-independent
errors are to be added in quadrature, and $E$ is to be measured in GeV.

\subsection{Tevatron}

To quantify the experimental sensitivity to the anomalous
couplings,  we first impose
acceptance cuts on the transverse momentum ($p_T$), 
pseudo-rapidity
($\eta$), and the separation in the azimuthal angle-pseudo rapidity plane
($\Delta R$) for the charged lepton and photon from the jets. We choose
the following ``basic'' acceptance cuts:
\begin{eqnarray}
& p_T^\ell> 15~{\rm GeV}, \qquad p_T^j> 20~{\rm GeV}, 
\qquad p_T^\gamma > 30~{\rm GeV}, \qquad
E_T^{miss} > 20 {\rm~GeV},  \nonumber\\
& |\eta^\ell|, |\eta^\gamma|, |\eta^j|<2.5, \qquad 
\Delta R_{\ell j}, \Delta R_{jj}, \Delta R_{\gamma j} > 0.4.  
\label{EQ:BASIC}
\end{eqnarray}
The high transverse momentum thresholds for the jets and the photon
are motivated by the hard $p_T$ spectrum from the heavy top decay.
With these basic cuts, the signal rate in Fig.~3(a) is reduced by about
40\% (with a maximal value of about 2.7~fb), while
the EW irreducible background of Eq.~(\ref{EQ:EW})
is about 30 fb, and the QCD process for
$\ell^\pm \nu j jj$ production in Eq.~(\ref{EQ:QCD}) is about 3000 fb. 
% e+-(mu+-) jjj (one of them has pT>30GeV).
It has been shown that the fake rate for a jet to a photon
($j\rightarrow \gamma$) at the Tevatron experiments \cite{cdfgm} can be
controlled down to a level of $10^{-4}$ for $p_T^\gamma>25$ GeV,
making it insignificant. Therefore, the EW background dominates and we
will consequently concentrate on that.

Figure~4(a) shows the reconstructed distributions for the top-antitop
invariant mass. Obviously, $M(t \bar t)$ for the signal has a
kinematical lower limit ($2m_t$ if no energy smearing is applied);
while the lower limit is significantly smaller for the background,
near the $W\gamma jj$ threshold. If we only accept events with
\begin{equation}
M(t\bar t) > 2m_t = 350 \ {\rm GeV},
\label{EQ:MTT}
\end{equation}
we expect to improve the signal-to-background ratio ($S/B$).
Also, due to the nature of top-quark two-body decay, the final
state jets and the photon have transverse momenta typically the
order of 
%$\frac{1}{2} m_t \, (1-M_W^2/m_t^2) \simeq 80$ GeV, 
$\frac{1}{2} m_t \simeq 80$ GeV, 
while all the jets and photon in the background events tend to be soft. 
To further discriminate the signal from the background, we define
two scalar sums of the transverse momenta:
$$
p^{}_T(cb) \equiv |\vec p^{c}_T|   + |\vec p^{b}_T|,  \qquad
p^{}_T(cb\gamma) \equiv p^{}_T(cb) + |\vec p^{\gamma}_T|.
$$
Figure~4(b) shows the distributions for these variables. We see indeed
that the signal spectra are much harder at the low end. On the other hand,
they are limited by the physical scale $2m_t$ for the signal 
at the high end, while they extend further for the background.
With these in mind, our choices of the cuts are
\begin{equation}
100 < p^{}_T(cb) <  300 \ {\rm GeV}, \qquad 150 < p^{}_T(cb\gamma) < 450 {\rm GeV}.
\label{EQ:PTSUM}
\end{equation}
With the additional cuts of Eqs.~(\ref{EQ:MTT}) and (\ref{EQ:PTSUM}),
the EW background is reduced to about 10 fb, while the signal may be as
large as 2.4~fb.

In Fig.~5 we show the reconstructed top-quark mass distributions
$M(c\gamma)$ and $M(bW)$, after making the additional cuts in
Eqs.~(\ref{EQ:MTT}) and (\ref{EQ:PTSUM}). We see from Fig.~5(a) that,
despite the improvement in $S/B$, the continuum background is still
above the $M(c\gamma)$ signal peak at
$m_t$. Further improvement can be made if we only study the events with
\begin{equation}
|M(c\gamma)-m_t| < 20 \ {\rm GeV}.
\label{EQ:MCG}
\end{equation}
The short-dashed histogram in
Fig.~5(b) shows how the background in the $M(bW)$ distribution is
reduced by the cut Eq.~(\ref{EQ:MCG}). We find that the signal peak
is nearly unaffected by this cut and is now above the background, so
that statistically significant effects may be observed near $m_t$
in the $M(bW)$ spectrum. If we examine the events in the mass range
\begin{equation}
|M(bW)-m_t| < 30 \ {\rm GeV},
\label{EQ:MBW}
\end{equation}
the signal observability should be nearly optimized. After the cut in
Eq.~(\ref{EQ:MBW}) we are left
with a maximal signal of 2.0 fb and an EW background 1.8 fb.
%The cross section for the signal after the cut in Eq.~19 
%(which further reduces the signal
%by only about 5\%) could be as large as about 2.3~fb.

\subsection{LHC}

At the LHC energy, due to the more complicated hadronic backgrounds,
we need to increase the transverse momentum thresholds, especially
for jets, direct photons and missing $E_T^{}$.
We adopt the ``basic'' acceptance cuts as
\begin{eqnarray}
& p_T^\ell> 20~{\rm GeV}, \qquad p_T^j> 35~{\rm GeV}, 
\qquad p_T^\gamma > 40~{\rm GeV}, \qquad
E_T^{miss} > 30 {\rm~GeV},  \nonumber\\
& |\eta^\ell|, |\eta^\gamma|, |\eta^j|<3, \qquad 
\Delta R_{\ell j}, \Delta R_{jj}, \Delta R_{\gamma j} > 0.4.  
\label{EQ:BASICLHC}
\end{eqnarray}
The signal rates given in Fig.~3(b) are reduced by about 60\%
with the cuts of Eq.~(\ref{EQ:BASICLHC}) (with a maximal value of about
0.22~pb), while the EW background is about 0.5 pb, and the ``faked'' QCD
background is about 70 pb. Assuming that we can effectively reject the QCD
background by photon identification to a level of $10^{-3}$ for $j \to
\gamma$ \cite{cmsatlas}, then the dominant background is again the
electroweak process. We see that after the basic cuts of 
Eq.~(\ref{EQ:BASICLHC}) the
S/B ratio is much closer to unity than in the Tevatron case.

Similar to the Tevatron study, we have constructed distributions for
the top-antitop invariant mass and scalar sums of transverse momenta (see
Fig.~6). We see that further background suppression can be achieved,
without much loss of signal, by the cuts of Eqs.~(\ref{EQ:MTT}) and
(\ref{EQ:PTSUM}). The signal integrated cross section after these cuts
may be as large as 200~fb, while the EW background rate is about 300~fb.
% QCD W3j=45 pb

As at the Tevatron, reconstructed top-quark mass distributions are also
crucial to non-ambiguously identify the signal. Figure~7 shows the
top-quark mass distributions reconstructed from the $c\gamma$ and $bW$ final
states after making the cuts in Eqs.~(\ref{EQ:MTT}) and (\ref{EQ:PTSUM}).
Once the final cuts in Eqs.~(\ref{EQ:MCG}) and (\ref{EQ:MBW}) are made,
the signal integrated cross section has a maximal value of 175 fb and
the EW background is 33 fb.  We see that the S/B ratio is
significantly better than at the Tevatron, primarily due to the huge
enhancement for the $t\bar t$ production at higher c.m. energies.

%\subsection{$t \bar t\to Wb c g$ at Hadron Colliders}
%defer this to Discussion

\subsection{$b$-tagging Effects}

The CDF collaboration were able to achieve 
about 50\% $b$-tagging efficiency \cite{btag} and
one hopes to reach about the same efficiency for
the LHC experiments \cite{cmsatlas}. The impurity
from the light quarks/gluons is assumed to be
1\%. We therefore can naively
expect to further suppress the backgrounds by a factor
of $1\%\times n_j$ (where $n_j$ is the number of jets) 
at a cost of signal reduction of 50\%. 

One has to be cautious in making this statement since
there is direct production of $b\bar b$ among the other
light quark/gluon jets in the background events. We
have explicitly calculated the $W\gamma b \bar b$
production rate and found the cross section is
about 0.12 fb at the Tevatron and 1.1 fb at the LHC
with the basic cuts plus Eqs.~(\ref{EQ:MTT}) and (\ref{EQ:PTSUM}).
This implies that the misidentified $b$ from light
quarks/gluons is indeed the major background source.
Consequently, we will not carry out the calculation
for the $b\bar b$ fraction in $Wjjj$ events, since this
channel is always smaller to begin with.

Table~\ref{T:ONE} lists the signal rate
(for $\kappa_g=0,\kappa_\gamma=0.16$),
EW and QCD backgrounds at the Tevatron and the LHC.
Comparison is made for the results with only cuts of
Eq.~(\ref{EQ:BASIC}) for Tevatron or Eq.~(\ref{EQ:BASICLHC}) 
for LHC and Eqs.~(\ref{EQ:MTT})and (\ref{EQ:PTSUM}),
and those plus $b$-tagging. 
We indeed obtain significant improvement for the signal/background
ratios, although the signal rate at the Tevatron is very limited.

%( will need a Table 2 here for $Wb c-gluon$ case. )

\section{Discussions and Summary}

To estimate the sensitivity to the anomalous couplings for
a given accumulated luminosity, we first establish the 
approximate cross section formula for the $t\to c\gamma$ case
in terms of the anomalous couplings:
\begin{equation}
\sigma = \frac{\kappa_\gamma^2} {(1+1.04\kappa_g^2)^2} 
\ \sigma(\kappa_\gamma=1,\kappa_g=0)
\label{EQ:XTCGM}
\end{equation}
where $\sigma(\kappa_\gamma=1,\kappa_g=0)$ is calculated for the
appropriate set of cuts, and the couplings are subject to the
constraints in Fig.~1.
The cross section $\sigma(\kappa_\gamma=1,\kappa_g=0)$ is
80~fb and 6.8~pb at the Tevatron and LHC, respectively, after
the full set of cuts is made.
We start with Poisson
statistics since the signal rate is often small,
especially at the Tevatron energy.
%Figure~\ref{FIG:SENSE}
Figure~8 presents the anomalous coupling $\kappa_\gamma$ versus
the luminosity needed to probe $\kappa_\gamma$ at 95\% Confidence
Level (CL).\footnote{In our approach, a 95\% CL in Poisson statistics roughly
corresponds to $S/\sqrt{S+B}=3$ in Gaussian statistics when the number
of events is large.}
The solid curves presented in each panel are for $\kappa_g=0$
and the dashed for $|\kappa_g|=0.9$. The upper curves for each case
are those with kinematical cuts only, while the lower ones are that
including $b$-tagging improvement. 
We see that at the Tevatron energy, minimal luminosity of about 
5 fb$^{-1}$ is needed in order to probe $\kappa_\gamma$ near the
current low energy constraints at 95\% CL. At the LHC with 100 fb$^{-1}$,
one expects to improve the sensitivity to $\kappa_\gamma$ by
more than one order of magnitude. For instance, for $\kappa_g=0$
and with $b$-tagging, the sensitivity at the Tevatron reaches
\begin{eqnarray}
{\rm for \; 5 \; fb}^{-1}: \; 
|\kappa_\gamma|&\sim&0.16 \quad {\rm and} \quad
BR(t \rightarrow c \gamma) \sim 1.3\times 10^{-3} \nonumber\\
{\rm  10 \; fb}^{-1}: \; 
|\kappa_\gamma|&\sim&0.12 \quad {\rm and} \quad
BR(t \rightarrow c \gamma) \sim 7\times10^{-4} \nonumber\\
{\rm 30 \; fb}^{-1}: \; 
|\kappa_\gamma|&\sim&0.08 \quad {\rm and} \quad
BR(t \rightarrow c \gamma) \sim 3\times10^{-4},
\label{EQ:KAPPATEV}
\end{eqnarray}
and at the LHC
\begin{eqnarray}
{\rm for \; 10 \; fb}^{-1}: \; 
|\kappa_\gamma|&\sim&0.02 \quad {\rm and} \quad
BR(t \rightarrow c \gamma) \sim 2\times10^{-5} \nonumber\\
{\rm 100 \; fb}^{-1}: \; 
|\kappa_\gamma|&\sim&0.01 \quad {\rm and} \quad
BR(t \rightarrow c \gamma) \sim 5\times10^{-6},
\label{EQ:KAPPALHC}
\end{eqnarray}
Alternatively, setting $\kappa_\gamma=1$
and letting $\Lambda$ vary (see Eq.~(\ref{Leff})), 
we can view this as a probe of the
scale of new physics up to $\Lambda/\kappa_\gamma \sim 10$~TeV
at the Tevatron and about 100 TeV at the LHC.

If $\kappa_g\ne0$, there of course is also the possibility of $t
\rightarrow c g$ decay, which is only mildly constrained by current
Tevatron data. Although the hadronic nature of the decay might make the
detection of this mode difficult, the jets resulting from this decay
will have large transverse momentum, and may allow one to distinguish
the signal from the background. There is also the possibility of
single top production in association with a charm quark \cite{tcg}.
The phenomenology of the $\bar t c g$ coupling at hadron colliders is
currently under study \cite{HHWYZ}.

\section{Acknowledgments}

We would like to thank Liangxin Li for collaboration in the
calculation of the $b\rightarrow s\gamma$ limits during an early stage of 
this work. We thank Roberto Peccei for discussions. BLY would like to
thank Hai-Yang Cheng and Shih-Chang Lee for warm hospitality extended
to him during his sabbatical leave at the Institute of physics,
Academia Sinica, Taipei, when part of the work was performed. He would
also like to thank the National Science Council for financial support.
XZ thanks G. Buchalla for discussions.
This work was supported in part by  the U.S.~Department 
of Energy under Contracts  DE-FG03-91ER40674 (T. Han) and DE-FG02-94ER40817 
(K.Whisnant, B.-L. Young and X.Zhang).  
  
%%%%%%%%% tables

\begin{table}[htb]
%\begin{table}[h]
\centering
\caption[]{Cross sections in units of fb for the 
$t\bar t \rightarrow \ell^\pm \nu bc\gamma$ signal 
($\kappa_g=0,\kappa_\gamma=0.16$),
and EW and QCD backgrounds at the Tevatron and the LHC.
Comparison is made for the results with only cuts 
Eqs.~(\ref{EQ:BASIC} or \ref{EQ:BASICLHC}) and
(\ref{EQ:MTT}-\ref{EQ:PTSUM})
and those plus $b$-tagging. 50\% tagging efficiency and 1\% impurity are
assumed \cite{btag}. }
\begin{tabular}{|c|c|c|c|} 
(a). Tevatron & signal $t \bar t \rightarrow \ell^\pm \nu bc\gamma$~~~ & 
$W\gamma jj \to \ell^\pm \nu \gamma jj$~~~  & 
$Wjjj \to \ell^\pm \nu \gamma jj$~~~ \\ \hline
cuts only & 2.4  &  10 &  0.66 \\ \hline
plus $b$-tag & 1.2 & 0.3 & 0.01\\ \hline\hline
(b). LHC & signal $t \bar t \rightarrow \ell^\pm \nu bc\gamma$~~~ & 
$W\gamma jj \to \ell^\pm \nu \gamma jj$~~~  & 
$Wjjj \to \ell^\pm \nu \gamma jj$~~~ \\ \hline
cuts only &  200 &  300 &  45 \\ \hline
plus $b$-tag & 100 &  7 &  1 \\
\end{tabular}
\label{T:ONE}
\end{table}
%

%%%%%%%%%%%%%%%%%%

%
\vfill
\eject
\centerline{FIGURE CAPTIONS}

FIG.~1 Constraints on anomalous couplings $\kappa_g$ and
$\kappa_\gamma$ from $BR(b \rightarrow s\gamma)$ (the two diagonal
bands in solid lines) and $BR(t \rightarrow bW)$ (dashed horizontal lines).
$\Lambda=1$~TeV is assumed throughout the paper.

\bigskip
FIG.~2 (a) Branching ratios for (a) $t \rightarrow c \gamma$ versus
$|\kappa_\gamma|$ for $|\kappa_g|=0$ (solid line) and $|\kappa_g|= 0.9$
(dashed line), and (b) $t \rightarrow c g$ versus $|\kappa_g|$.
The curves stop at the maximum allowed value of the abscissa in each case.

\bigskip
FIG.~3 Production cross section for 
$t{\bar t} \rightarrow bWc\gamma \rightarrow b\ell\nu c\gamma$
at (a) the Tevatron ($p{\bar p}$)
with $\sqrt{s}=2$~TeV, and
(b) the LHC ($p p$) with $\sqrt{s}=14$~TeV. The top-quark mass
$m_t=175$~GeV is used throughout the paper.
The cross sections are plotted versus $|\kappa_\gamma|$ for
$|\kappa_g|=0$ (solid line) and $|\kappa_g|=0.9$ (dashed line). The curves
stop at the maximum allowed value of $|\kappa_\gamma|$ in each case.

\bigskip
FIG.~4 (a) Top-antitop invariant mass distributions for the signal
$t{\bar t} \rightarrow Wbc\gamma$ (solid histogram) and background
$p{\bar p} \rightarrow Wjj\gamma$ (dashed histogram) at the Tevatron
with $\sqrt{s} =2$~TeV, assuming the basic acceptance cuts defined in
Eq.~(\ref{EQ:BASIC}) in the text.
(b) Distributions for the transverse momenta scalar sums $p_T(bc)$
and $p_T(bc\gamma)$ for the signal (solid histograms) and background
(dashed histograms). The signal histograms assume the maximal allowed
value of $\kappa_\gamma$ with $\kappa_g=0$.

\bigskip
FIG.~5 Reconstructed top-quark mass distributions for (a) $M(c\gamma)$
and (b) $M(bW)$ at the Tevatron with $\sqrt{s} =2$~TeV. In (a), the solid
histogram shows the signal $t{\bar t} \rightarrow Wbc\gamma$ and the
short dashed histogram shows the background $p{\bar p} \rightarrow Wjj\gamma$
after applying basic cuts and cuts on the top-antitop invariant mass and the
transverse momenta scalar sums $p_T(bc)$ and $p_T(bc\gamma)$, while in
(b) the same curves are used to show the signal and background after
making the further cut on $M(c\gamma)$. The long dashed histogram in
(b) shows the $M(bW)$ distribution of the background before $M(c\gamma)$
cut is made (the effect of the $M(c\gamma)$ cut on the signal is minimal).
The signal histograms assume the maximal allowed value of
$\kappa_\gamma$ with $\kappa_g=0$.

\bigskip
FIG.~6 (a) Top-antitop invariant mass distributions for the signal
$t{\bar t} \rightarrow Wbc\gamma$ (solid histogram) and background
$p{\bar p} \rightarrow Wjj\gamma$ (dashed histogram) at the LHC
with $\sqrt{s} =14$~TeV, assuming the basic acceptance cuts defined in
Eq.~(\ref{EQ:BASICLHC}) in the text.
(b) Distributions for the transverse momenta scalar sums $p_T(bc)$
and $p_T(bc\gamma)$ for the signal (solid histograms) and background
(dashed histograms). The signal histograms assume the maximal allowed
value of $\kappa_\gamma$ with $\kappa_g=0$.

\bigskip
FIG.~7 Reconstructed top-quark mass distributions for (a) $M(c\gamma)$
and (b) $M(bW)$ at the LHC with $\sqrt{s} =14$~TeV. In (a), the solid
histogram shows the signal $t{\bar t} \rightarrow Wbc\gamma$ and the
short dashed histogram shows the background $p{\bar p} \rightarrow Wjj\gamma$
after applying basic cuts and cuts on the top-antitop invariant mass and the
transverse momenta scalar sums $p_T(bc)$ and $p_T(bc\gamma)$, while in
(b) the same curves are used to show the signal and background after
making the further cut on $M(c\gamma)$. The long dashed histogram in
(b) shows the $M(bW)$ distribution of the background before $M(c\gamma)$
cut is made (the effect of the $M(c\gamma)$ cut on the signal is minimal).
The signal histograms assume the maximal allowed value of
$\kappa_\gamma$ with $\kappa_g=0$.

\bigskip
FIG.~8 95\% CL sensitivity to $\kappa_\gamma$ vs. integrated luminosity for
(a) the Tevatron with $\sqrt{s}=2$~TeV and (b) the LHC with
$\sqrt{s}=14$~TeV. We consider the limits for $\kappa_g=0$
(solid lines) and $|\kappa_g|=0.9$ (dashed lines). The upper curves
in each case correspond to the limit obtained when comparing the signal
and background after making all the cuts, while the lower curves
correspond to the limits obtained when using $b$-tagging with the cuts in
Eqs.~(\ref{EQ:BASIC} or \ref{EQ:BASICLHC}) and
(\ref{EQ:MTT}-\ref{EQ:PTSUM}). The
curves cut off at the maximal allowed value of $\kappa_\gamma$ in each case.

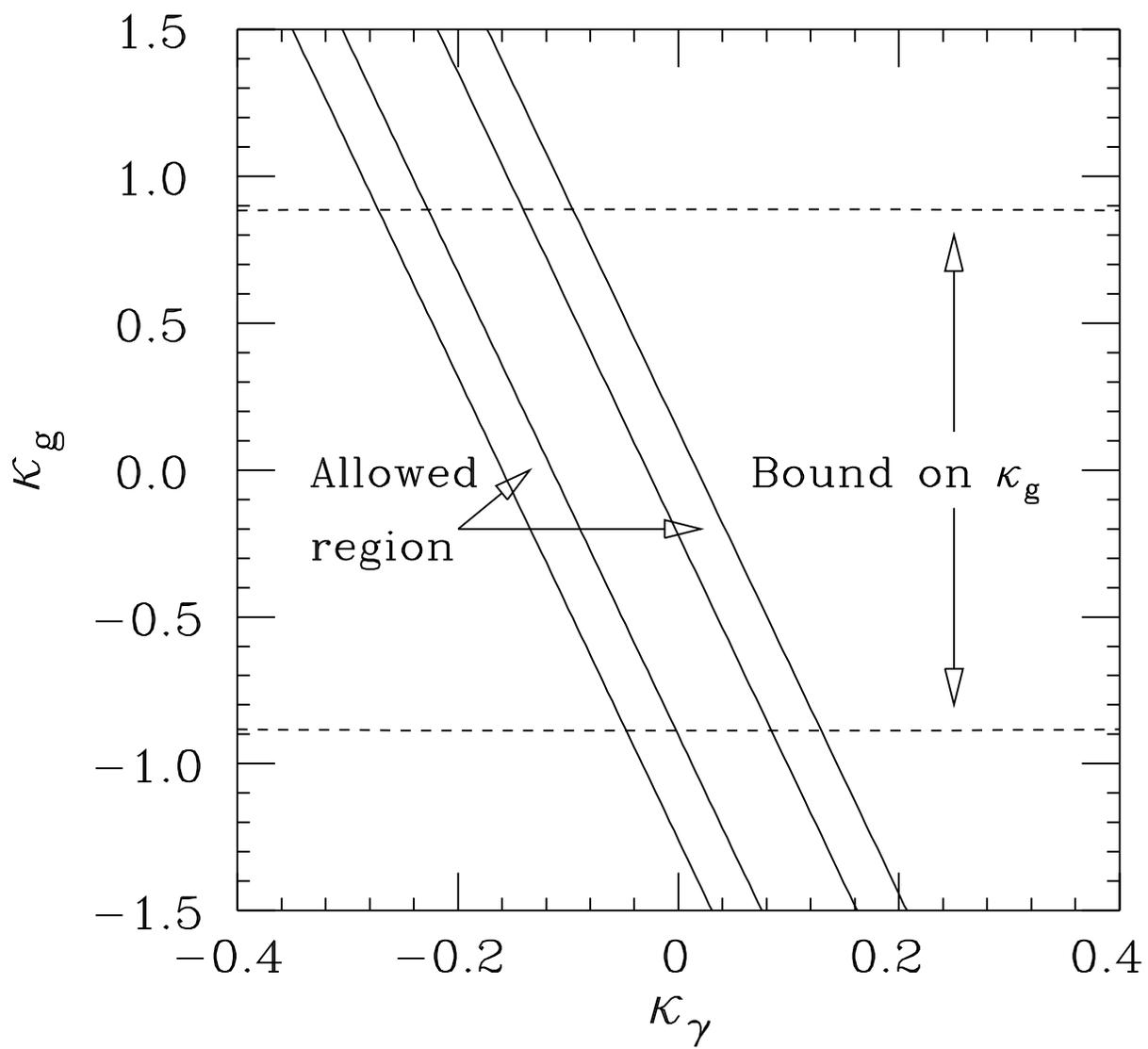
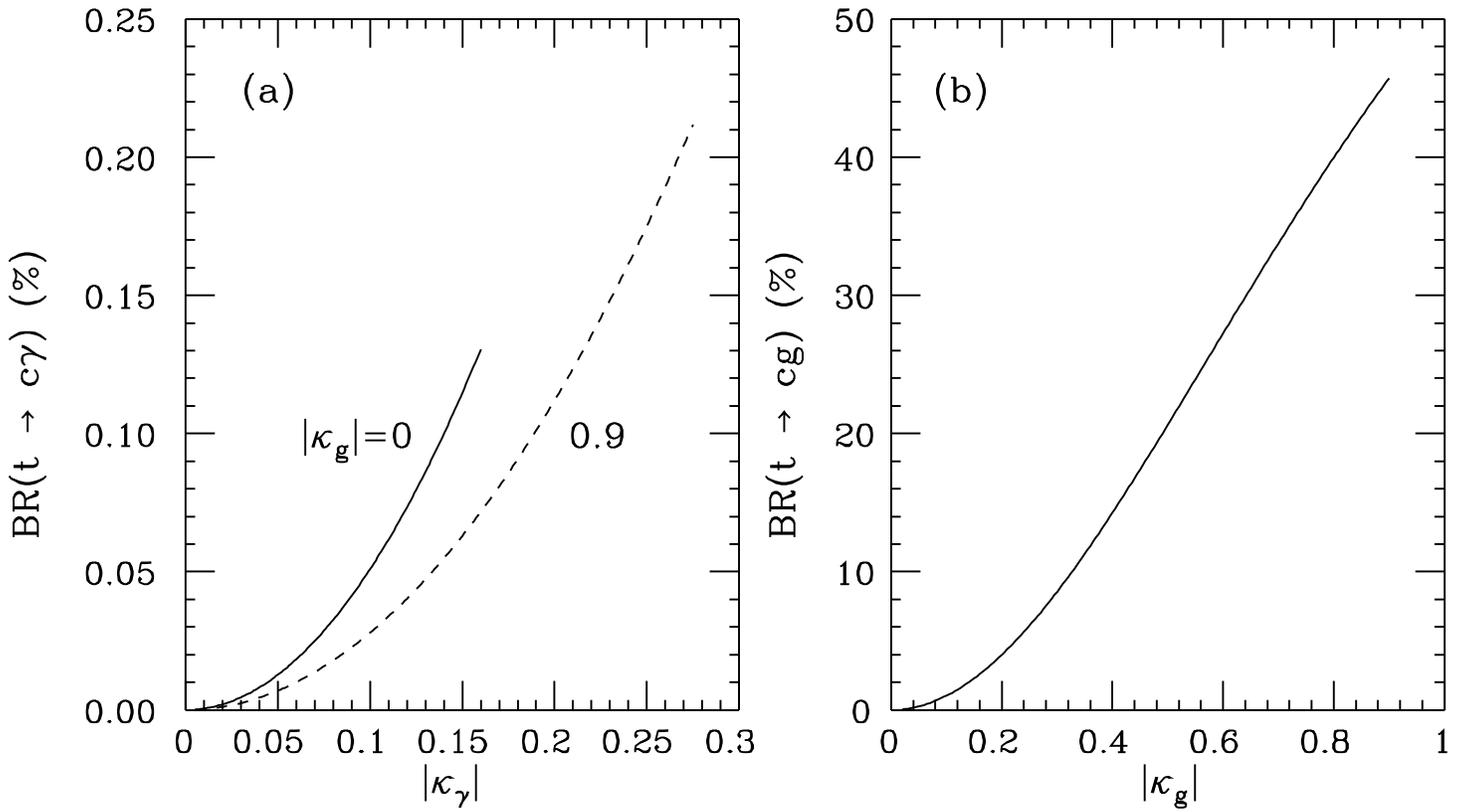
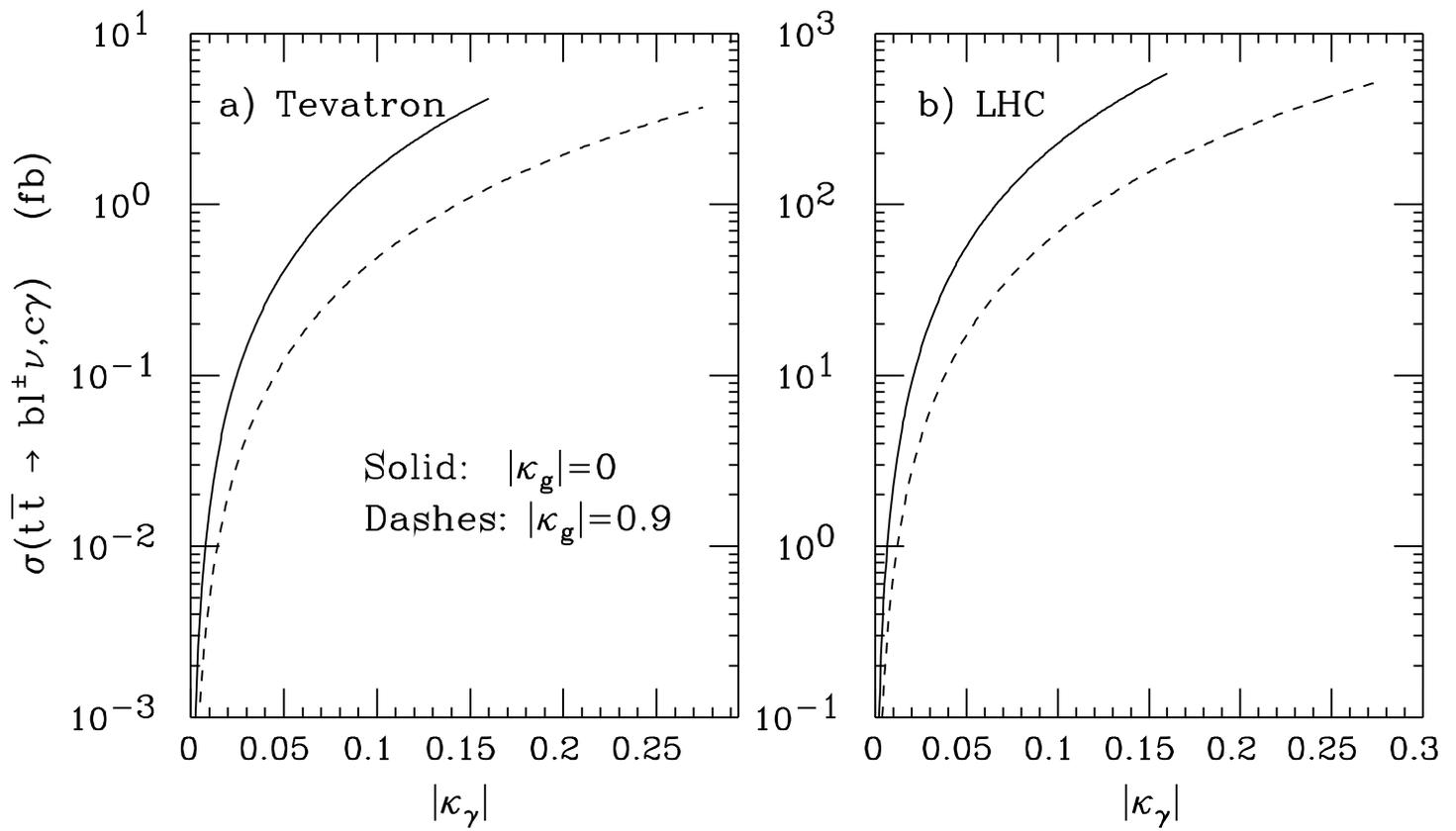
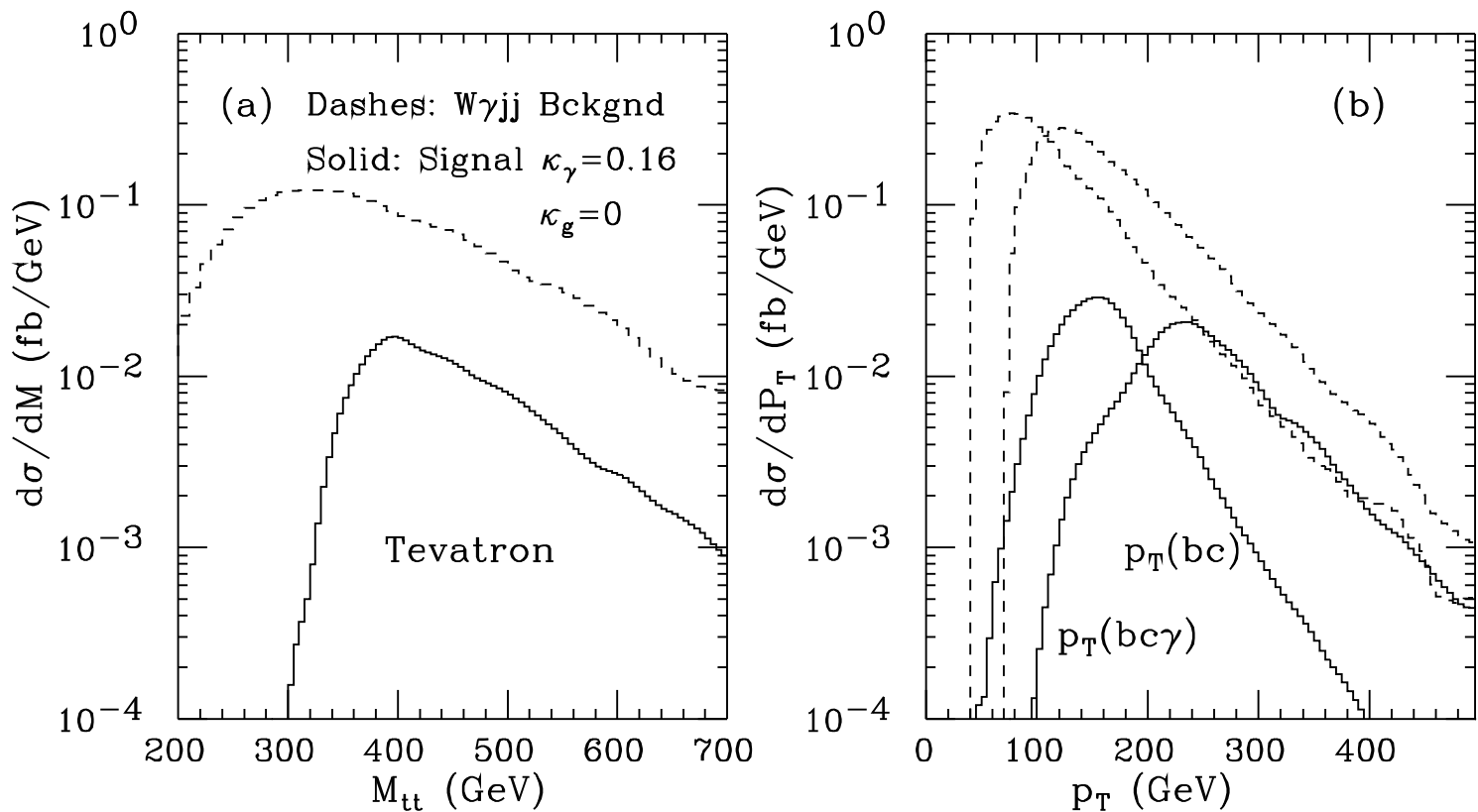
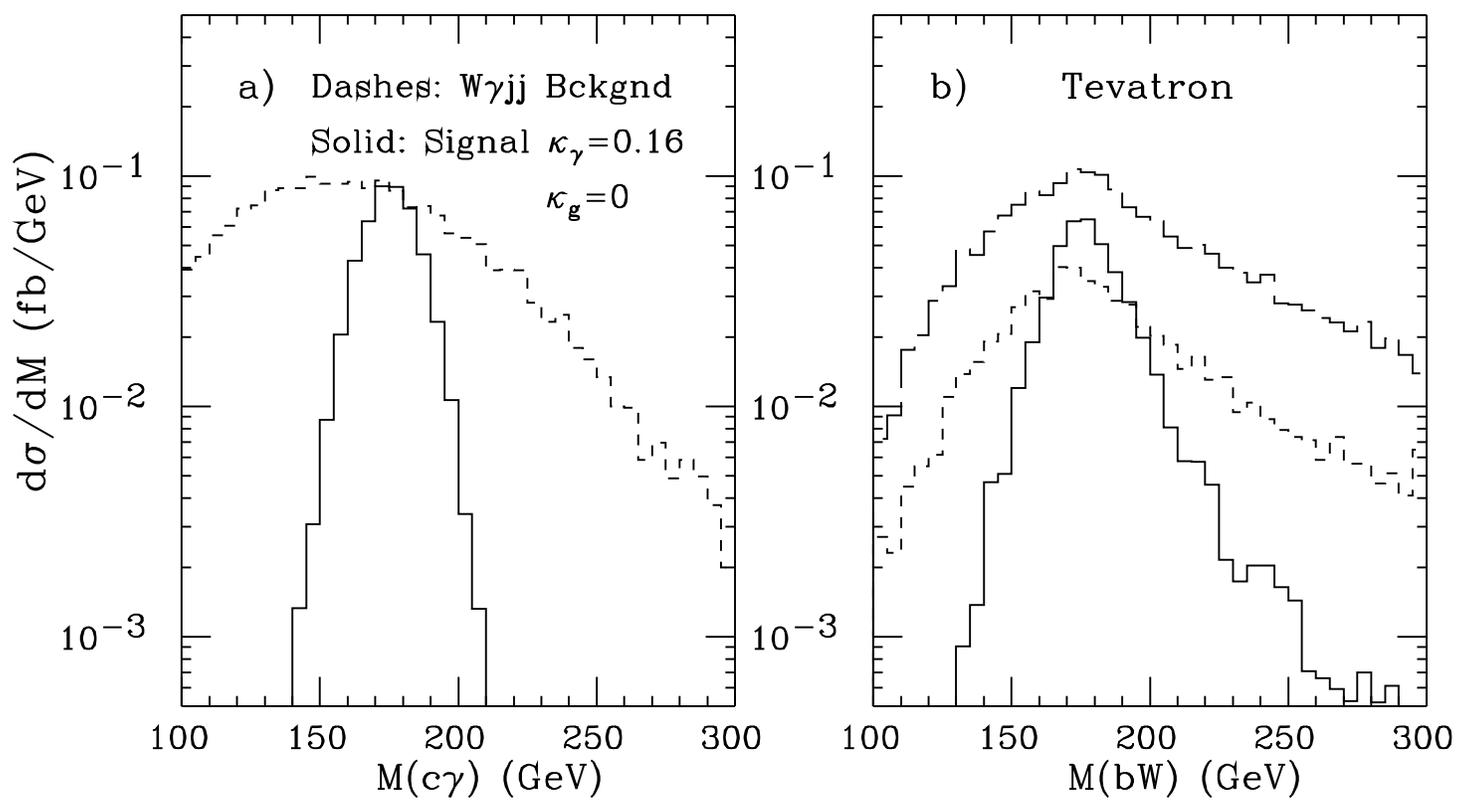
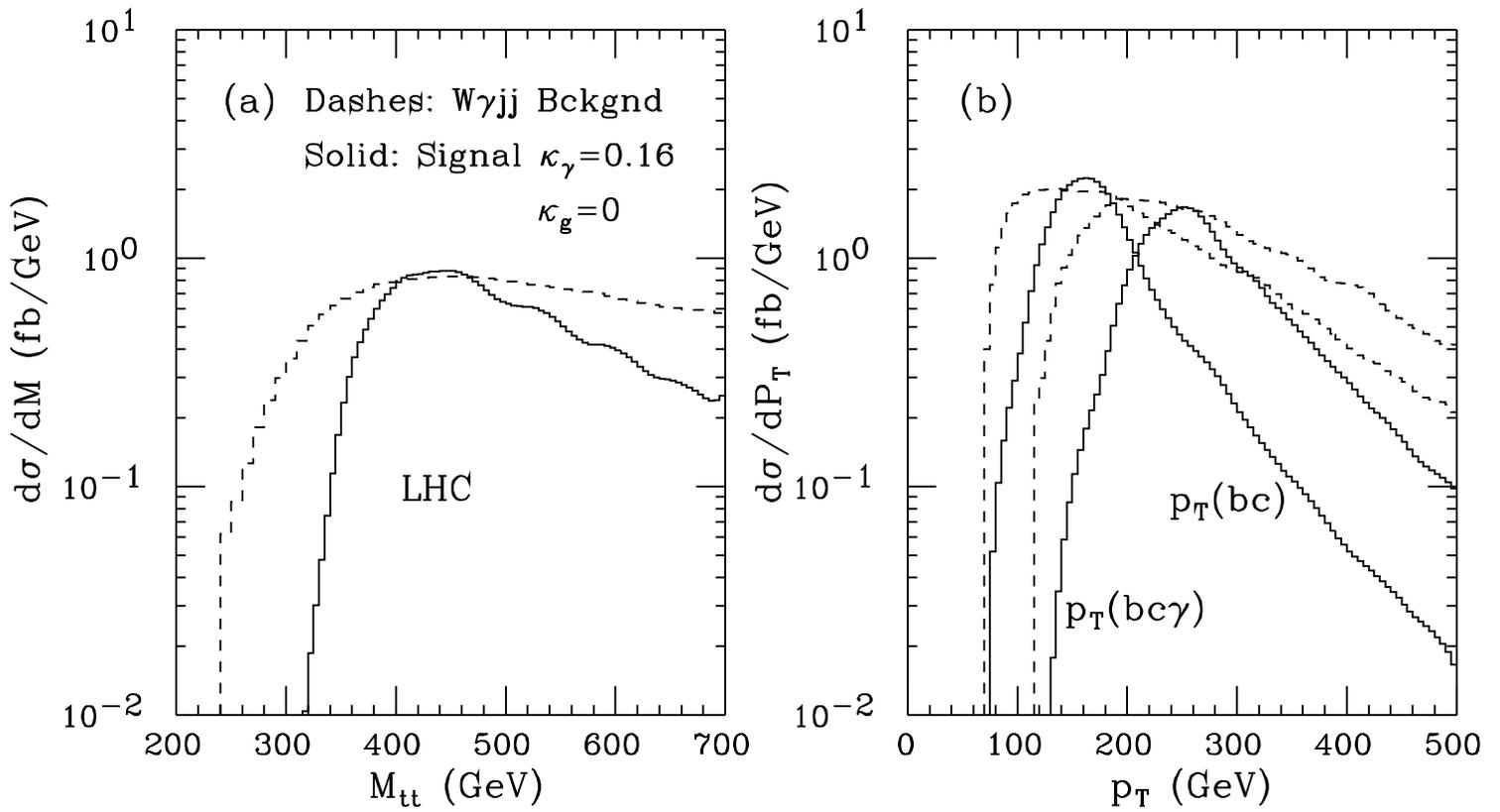
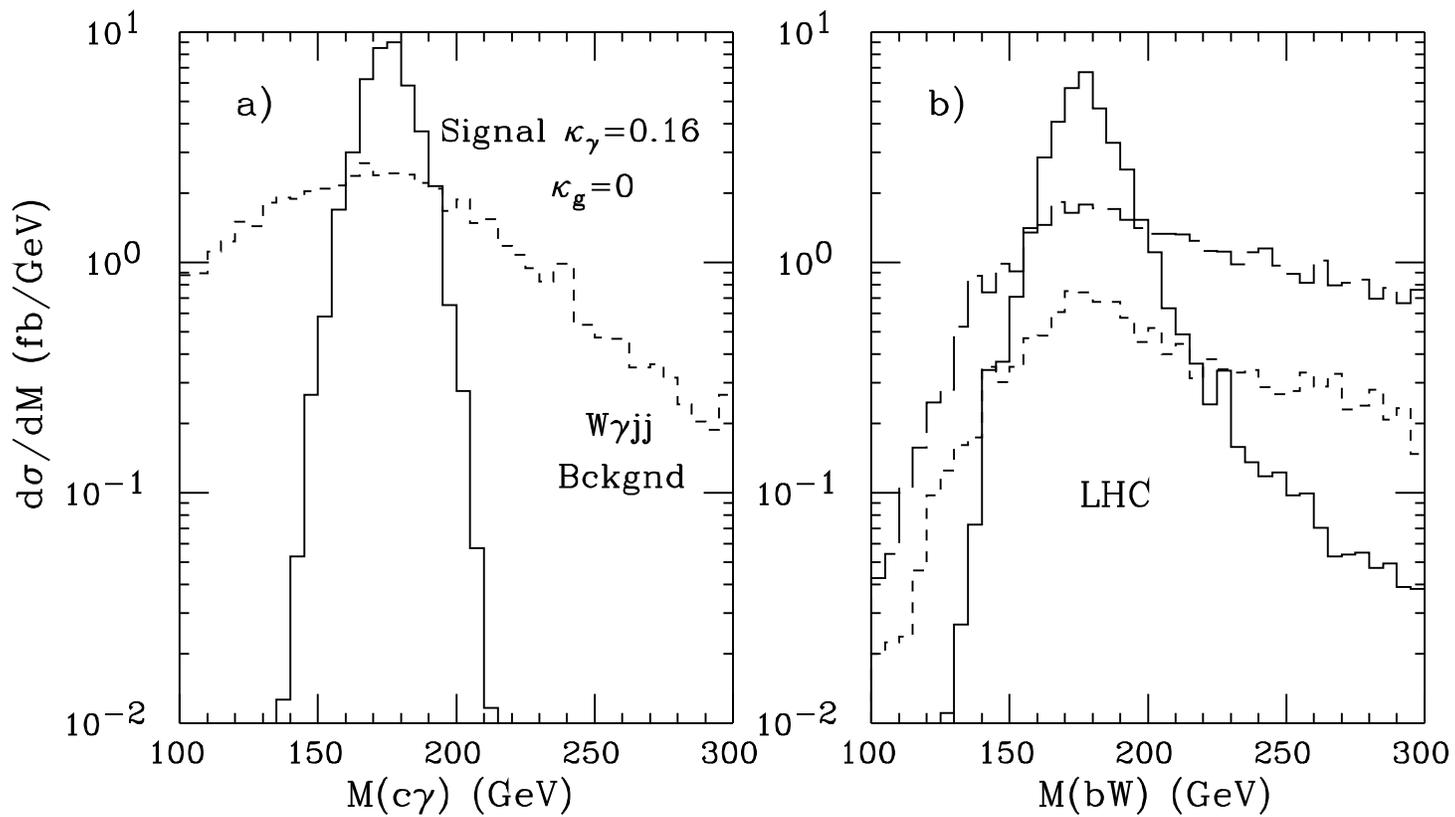
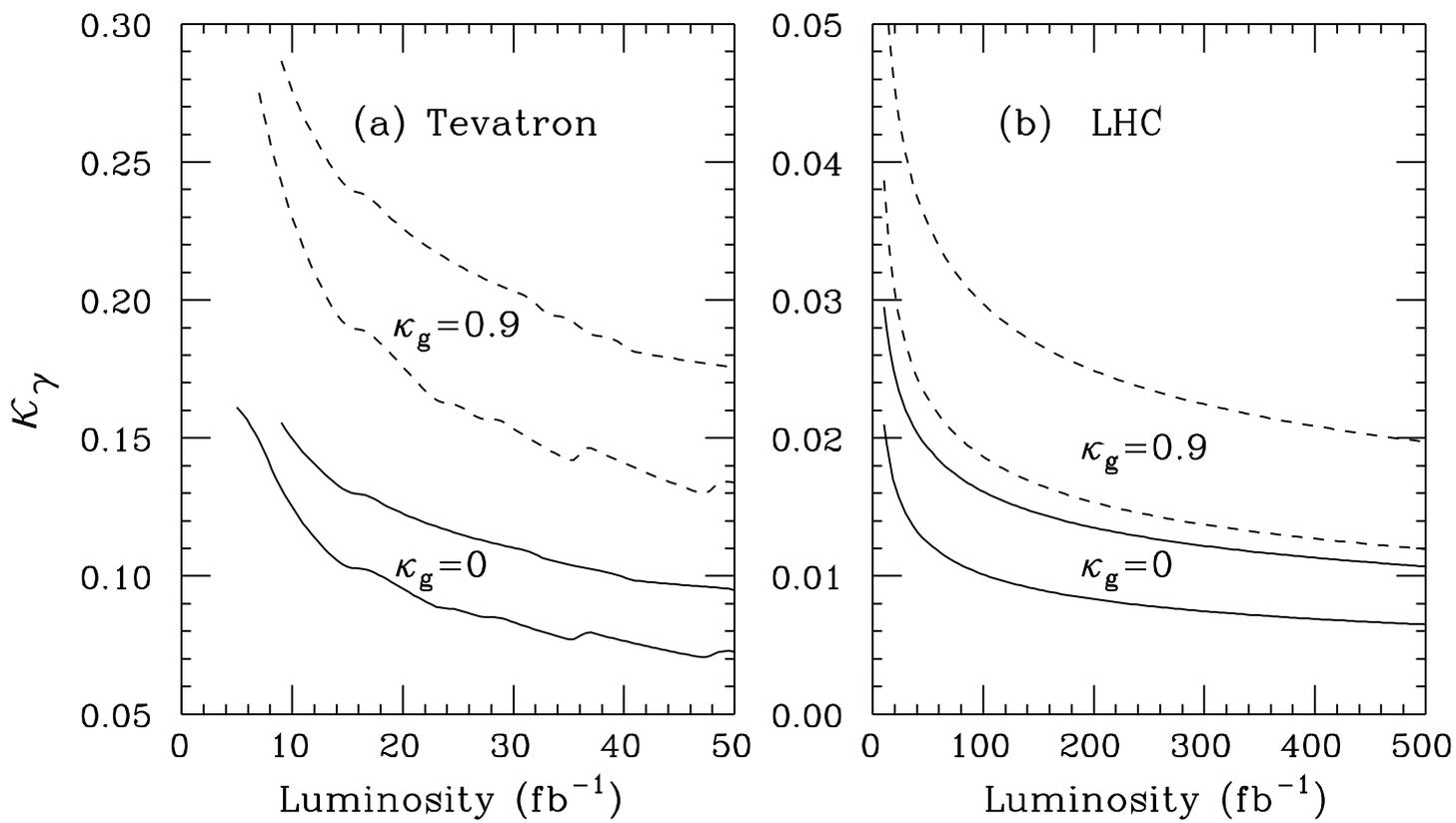
\end{document}